\documentclass[twocolumn,superscriptaddress,showpacs,amsmath,amssymb,prl,floatfix]{revtex4}
\usepackage{graphicx}
\usepackage{hyperref}
\begin{document}
\author{A.\ L\"auchli}
\affiliation{
  Institut Romand de Recherche Num\'erique en Physique des Mat\'eriaux (IRRMA),
  PPH-Ecublens, CH-1015 Lausanne
}
\author{J.C. Domenge}
\author{C. Lhuillier}
\author{P. Sindzingre}
\affiliation{
  Laboratoire de Physique Th\'eorique des
  Liquides, Universit\'e P. et M. Curie, case 121, 4 Place Jussieu, 75252
  Paris Cedex. UMR 7600 of CNRS
}
\author{M.\ Troyer}
\affiliation{
  Institut f\"ur Theoretische Physik, ETH H\"onggerberg, CH-8093 Z\"urich, Switzerland
}
\date{\today}
\title{Two-Step Restoration of SU(2) Symmetry in a Frustrated Ring Exchange Magnet}

\begin{abstract}
  We demonstrate the existence of a spin-nematic, moment-free phase in a quantum four-spin
  ring exchange model on the square lattice. This unusual quantum state is created by the interplay
  of frustration and quantum fluctuations which lead to a partial restoration of
  $SU(2)$ symmetry when going from a four-sublattice orthogonal biaxial N\'eel order to this
  exotic uniaxial magnet. A further increase of frustration drives a transition to a
  fully gapped $SU(2)$ symmetric valence bond crystal.
\end{abstract}
\maketitle

Broken symmetries are one of the central paradigms of magnetic ordering, and most
antiferromagnetic systems are in N\'eel phases at low temperatures, characterized by
a vectorial order parameter: their sublattice magnetic moment.
This order parameter breaks the rotational $SU(2)$ symmetry
of the Hamiltonian and is accompanied by gapless excitations,
the Goldstone modes of the broken symmetry \cite{auer94}.
In low-dimensional systems of Heisenberg spins frustrating couplings
can drive transitions to gapful $SU(2)$ symmetric quantum states,
where the building blocks are local singlets
(in the simplest examples, pairs of spin-1/2 in short-range singlet states).
These quantum gapped phases may have long-ranged singlet order and break spatial
symmetries of the lattice, called valence bond crystals (VBC) in the following.
An even more exotic groundstate, a coherent superposition of all lattice-coverings
by local singlets, a state known as a resonating valence bond (RVB) spin liquid, could
also be found~\cite{ml03}.

Quantum Phase Transitions from N\'eel ordered phases to quantum gapped phases have
been studied for a long time as prototypical examples of quantum phase transitions.
The nature of the symmetry breaking N\'eel phase plays an important role in these
scenarios and seems determinant as to whether the adjacent gapped phase will be a
VBC or an RVB phase~\cite{s03}. It was recently shown that the transition from the
standard collinear $(\pi,\pi)$ N\'eel state to a Valence Bond Crystal phase can
actually be an exotic quantum critical point with deconfined excitations~\cite{svbsf04}.

The well-known $(\pi,\pi)$ N\'eel state is a uniaxial magnet, with
two gapless Goldstone modes, in which $SU(2)$ is partially broken
to $U(1)$. More complete $SU(2)$ breaking schemes do exist, for
example in noncollinear magnets with more than two ferromagnetic
sublattices or more generally in helicoidal antiferromagnets. In
these systems, the order parameter can be described as biaxial (or
as a top), the $SU(2)$ symmetry is completely broken, and there
are three Goldstone modes. Chandra and Coleman suggested that in
such situations the restoration of the full $SU(2)$ symmetry due
to the interplay of quantum fluctuations and frustration could
possibly occur in two steps: from a biaxial magnet via an
intermediate uniaxial {\em spin nematic} magnet -- still with
gapless excitations -- to a fully gapped paramagnetic phase
without $SU(2)$ symmetry breaking~\cite{cc91,c91}. This speculated
spin-nematic phase -- first introduced by Andreev and
Grishchuk~\cite{ag84} -- has {\em no} net magnetic moment, but
nevertheless breaks the $SU(2)$ symmetry, as the individual spins
- albeit disordered - remain correlated in a plane. Up to now this
conjectured two-step scenario misses a concrete realization, as it
has not been found in the models originally proposed, e.g. the
$J_1-J_3$ model on the square lattice or various models on the
kagome lattice.

In this Letter we present the phase diagram of a frustrated
four-spin ring exchange model for $S=1/2$ on the square lattice.
We show that this model has a four-sublattice
orthogonal N\'eel groundstate with a biaxial order parameter.
This groundstate maximizes the square of the vectorial chirality, and 
appears as a natural "unfrustrated" starting point in the following.
Increasing the frustration by a change in the nearest neighbor antiferromagnetic
coupling, provides a first realization of the mechanism speculated by 
Chandra and Coleman~\cite{cc91}. Based on the analysis of spin-resolved spectra,
the transition from a biaxial to a uniaxial magnet - accompanied by the reduction 
from three Goldstone modes in the biaxial magnet to only two Goldstone
modes in the uniaxial one - is highlighted. Upon a further increase in the
frustration all spin excitations become gapped and the system enters a (staggered) 
Valence Bond Crystal phase.

\begin{figure}
  \centerline{\includegraphics[width=0.6\linewidth]{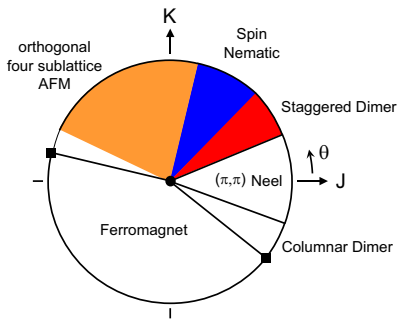}}
  \caption{(Color online)
    Schematic phase diagram of the cyclic four-spin exchange
    Hamiltonian~(\protect{\ref{eqn:Hamiltonian}}) on the square lattice
    as a function of the parameter $\theta$.
    \label{fig:PhaseDiagram}
  }
\end{figure}

The model Hamiltonian with spin $S=1/2$ reads:
\begin{eqnarray}\label{eqn:Hamiltonian}
  H=
  J \sum_{<i,j>}
  {\bf S}_i\cdot{\bf S}_j
  + K \sum_{
    [i,j,k,l]
  } \left( P_{i\ldots l}^{}+P_{i\ldots l}^{-1}\right).
\end{eqnarray}
$ P_{i\ldots l}$ is the cyclic permutation of the 4 spins sitting on
a square plaquette, and the two sums run respectively over the nearest neighbor bonds
and the plaquettes of the square lattice. We parametrize
the couplings by $J=\cos \theta$ and $K=\sin \theta$. The explicit expression of the
4-spin permutation operators in terms of spin-1/2 operators can be found
in Ref.~\cite{rhd83}.

Interest in the present model has arisen 15 years ago in an early
attempt to describe the magnetism of the
cuprates\cite{rd89,cgb92}. It has been shown very recently to be
relevant to describe the high-energy spin waves dispersion of
La$_2$CuO$_4$~\cite{c01}. In this material, the four-spin ring
exchange term is significant, but still rather small compared to
the usual Heisenberg term \cite{NoteCyclic}.
Here we study this model in a broader range of parameters.
Recent studies of the Hamiltonian (\ref{eqn:Hamiltonian}) on a two-leg ladder
have provided evidence for a number of unconventional phases \cite{lst03,hmx03}.
Of particular relevance for the present work is the appearance of a
short range ordered, gapped phase with dominant {\em vector chirality} correlations
in a large region of the phase diagram.

Our results on the overall phase diagram obtained by large scale exact
diagonalizations of systems up to $N=40$ spins are summarized in
Fig.~\ref{fig:PhaseDiagram}.
Adjacent to the well established N\'eel phase around $J=1,\ K=0$, we
find two VBC phases, a staggered dimer phase for positive $K$ and a
columnar dimer phase for negative $K$. For negative $K$ we further find
a large ferromagnetic region. These phases for $K<0$ are analogous
to those observed in a related $XY$-like model, amenable to Quantum
Monte Carlo simulations~\cite{sdss02}. The emphasis of the present
paper is on three of the phases found for $K>0$: the coplanar
orthogonal antiferromagnet, the spin-nematic state and the staggered dimer
VBC. A detailed discussion of the whole phase diagram will be presented
in a forthcoming article \cite{ldslt05}.

\begin{figure}
  \centerline{\includegraphics[width=0.95\linewidth]{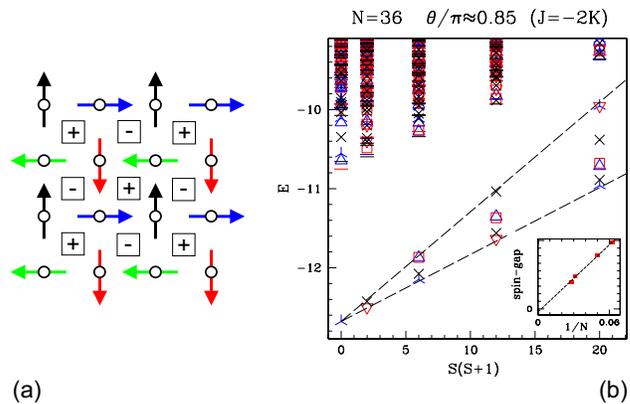}}
  \caption{
    (Color online)
    (a) The orthogonal N\'eel state. It has a coplanar, four-sublattice
    structure. The vector chirality on a plaquette possesses a $(\pi,\pi)$ structure.
    (b) Tower of states in the orthogonal four-sublattice N\'eel state
    for $J=-2,\ K=1$ $(\theta\approx 0.85\pi)$. The eigenstates between the dashed lines
    form the groundstate manifold of the orthogonal N\'eel state. Taking into account
    exact degeneracies of some of the states we find $2S+1$ low-lying levels
    in each spin-$S$ sector,
    consistently with a complete $SU(2)$-symmetry breaking order parameter.
    Inset: finite size scaling of the spin gap, indicating a vanishing
    spin gap in the thermodynamic limit.
    \label{fig:TowerOrthogonal}
  }
\end{figure}

{\it The orthogonal four-sublattice AFM --- } Let us start the
discussion with a state which has a simple classical interpretation.
The cyclic exchange term $K$ contains both four-spin couplings and two-spin exchange terms.
For $J=-2$ and $K=1$ ($\theta \approx 0.85\pi$),
the effective nearest neighbor
exchange coupling is vanishing \cite{rhd83}.
Classical and semiclassical reasonings then predict an orthogonal four-sublattice
AFM \cite{rd89,cgb92}, as shown in Fig.~\ref{fig:TowerOrthogonal}a).
It could be remarked that this state minimizes the  $SU(2)$ and lattice symmetric Hamiltonian:
${\cal H}'=-\sum_i{\bf {\cal C}}_i^2$, where $\cal C$ is the vectorial 
chirality on an elementary plaquette:
\begin{equation}
  \label{eqn:Chirality}
  {\cal C} = {\bf S}_1\wedge {\bf S}_2 + {\bf S}_2\wedge {\bf
    S}_3 +{\bf S}_3\wedge {\bf S}_4 + {\bf S}_4\wedge {\bf S}_1.
\end{equation}
Interestingly this Hamiltonian is
rather close to our model Eq.~(\ref{eqn:Hamiltonian}) in the region
$J=-2,\ K=1$.

\begin{table}
  \begin{tabular}{|ccc|c|c|c|}
    \hline
    \textbf{k} & ${\cal R}_{\pi/2}$ & $\sigma$ & (A) & (B) $N=8p$ & (B) $N=8p+4$ \\
    \hline
    $(0,0) $           & $1$  & $1$ & $\checkmark$ & $\checkmark$  & \\
    $(0,0) $           & $-1$ & $1$ & $\checkmark$ &   & $\checkmark$ \\
    $(\pi,\pi) $       & $1$  & $1$ & $\checkmark$ &   & $\checkmark$ \\
    $(\pi,\pi)$        & $-1$ & $1$ & $\checkmark$ & $\checkmark$ &  \\
    $(0,\pi),(\pi,0) $ & $0$  & $1$ & $\checkmark$ &   &  \\ 
    \hline
  \end{tabular}
  \caption{
    Irreducible representations of the square lattice space symmetry group
    appearing in the tower of states of the orthogonal N\'eel state {\bf (A)} and
    the spin-nematic state {\bf (B)}  respectively. \textbf{k} denotes the
    momentum, ${\cal R}_{\pi/2}$ the $90^\circ$ rotation around a site, and
    $\sigma$ the site-based reflection along the x or y axis. $N$ is the number
    of sites of the sample.
    \label{tab:SymSectors}}
\end{table}

We now turn to the $S=1/2$ fully quantum case, where the presence
of the orthogonal N\'eel state has not yet been shown explicitly.
We investigate the nature of the collective ground-state
with a spectroscopic method which relies on the analysis of symmetries and scaling
of the low lying levels of the exact spectra of the Hamiltonian~(\ref{eqn:Hamiltonian})
(called Quasi Degenerate Joint States (QDJS) in the following) \cite{a84,bllp94}.
These QDJS embody the dynamics and symmetry of the order parameter
of the orthogonal N\'eel state. As it is  a biaxial magnet, its
order parameter is a {\em quantum top}. The leading term of the free
dynamics of a quantum top is proportional to the square of the
total spin of the sample $S$: its effective spectrum
involves $(2S+1)$ distinct eigenstates in each $S$ sector,
with eigenvalues scaling as $S(S+1)/N$.

Fig.~\ref{fig:TowerOrthogonal}b) indeed displays such a tower of low lying levels
well separated from the other excitations. The symmetries of the QDJS (displayed in
Tab.~\ref{tab:SymSectors}) are those predicted by the ab initio symmetry analysis; 
three soft modes at $(0,\pi)$,  $(\pi,0)$, $(\pi,\pi)$ signal the full symmetry 
breaking of $SU(2)$. The finite size scaling of the QDJS is regular 
and as expected, the tower of states collapse to the ground-state as $1/N$ 
(Inset in Fig~\ref{fig:TowerOrthogonal}b) and \cite{ldslt05}).
Long wave-length quantum fluctuations, estimated in a spin wave approach, lead to a
reduction  of $\sim 30\%$ of the sublattice magnetization 
the thermodynamic limit. The real-space spin correlations as well as the vector 
chirality correlations are in perfect agreement with these results. Based on the analysis
of the exact spectra and finite size scaling of the orderparameters we believe that the
four-sublattice N\'eel phase is stable for 
$0.4\ \pi \lesssim \theta \lesssim 0.9\ \pi$.

\begin{figure}
  \centerline{\includegraphics[width=\linewidth]{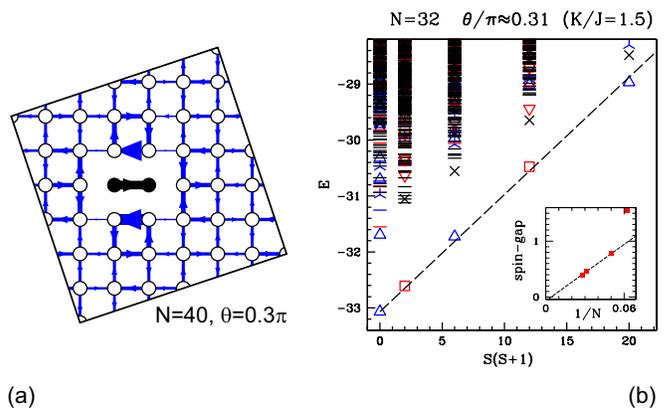}}
  \caption{
    (Color online)
    (a) Real space vector chirality correlations
    $\langle[{\bf S}_0\wedge {\bf S}_1]^z[{\bf S}_i\wedge {\bf S}_j]^z\rangle$
    for a $N=40$ sample in the spin-nematic phase at $\theta=0.3\pi$.
    The black bond denotes the oriented reference bond. The width of the lines
    is proportional to the correlation strength.
    (b) Tower of states in the spin-nematic state.
    Inset: finite size scaling of the spin gap, indicating a vanishing
    spin gap in the thermodynamic limit.
    \label{fig:TowerNematic}
  }
\end{figure}

{\em The spin-nematic phase ---}
Frustrating the four-sublattice orthogonal state by increasing $J$ induces a 
drastic modification of the low lying spectrum of Eq.~\ref{eqn:Hamiltonian}, 
which evolves towards the typical behavior of Fig.~\ref{fig:TowerNematic}b).
The $1/N$ finite size scaling of this tower of states proves that this phase 
breaks $SU(2)$ symmetry [Inset of Fig.~\ref{fig:TowerNematic}b)]. 
But the QDJS which display only one level in each $S$
sector, embed the dynamics of a rigid rotator: the magnet is a uniaxial magnet,
i.e. $SU(2)$ is only broken down to $U(1)$. 
One observes an enlargement of the spatial symmetry of the order parameter 
(see column (B) of Table~\ref{tab:SymSectors}), incompatible with a standard 
$(\pi,\pi)$ antiferromagnet, but consistent with a staggered long range order in
the vectorial chirality (\ref{eqn:Chirality}). This is confirmed by
the behavior of the correlations in the bond chirality
(defined as $\vec{\cal{V}}(i,j)=\langle{\bf S}_i\wedge {\bf S}_j\rangle$)
shown in Fig.~\ref{fig:TowerNematic}a).
On the other hand the finite size scaling
of the spin-spin correlations points to a wiping out of the sub-lattice
magnetization by long wave-length quantum fluctuations. Such a state is
therefore a $p$-spin-nematic state \cite{ag84,cc91,c91}, characterized by 
the absence of any sublattice magnetic moment $\langle {\bf S}_i\rangle=0$, 
and by the presence of a pseudo-vectorial order parameter $\vec{\cal{V}}(i,j)\neq 0$.

\begin{figure}
  \centerline{\includegraphics[width=0.6\linewidth]{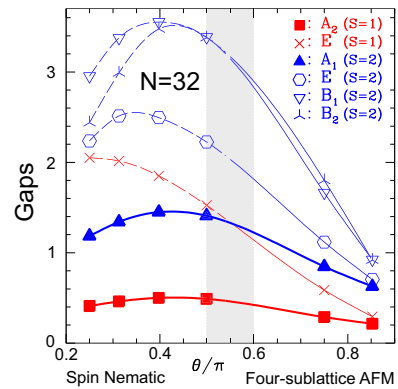}}
  \caption{
    (Color online)
    The evolution of the finite size spectral gaps within the QDJS of the orthogonal 
    N\'eel state on a $N=32$ sample.
    The bold lines denote levels which remain in the QDJS of the spin-nematic state.
    The other levels detach from the QDJS as $\theta \lesssim \pi/2$.
    \label{fig:TowerSplitting}
  }
\end{figure}

The partial restoration of the $SU(2)$ symmetry when going from
the four-sublattice orthogonal state to the nematic state can be
tracked by plotting the relative motion of the different 
symmetry-breaking levels within the tower
of QDJS while lowering $\theta$. The energy differences displayed
in Fig.~\ref{fig:TowerSplitting} show how all but {\em one} level
for each spin sector evaporate once $\theta/\pi\lesssim 0.5$.
Since the symmetry group of the orthogonal four-sublattice
antiferromagnet is contained in the symmetry group of the
spin-nematic state we might expect the transition between the two
states to be a continous quantum phase transition, although this
remains an open problem.

The finite size scaling of the order parameter indicates that the
phase should at least exist in the range of parameters $0.25 \lesssim \theta /\pi \lesssim 0.4 $. 
The accuracy in the determination of the boundaries cannot be made better on the basis of
exact diagonalizations.

{\em The staggered dimer VBC phase ---}
\begin{figure}
  \centerline{\includegraphics[width=0.9\linewidth]{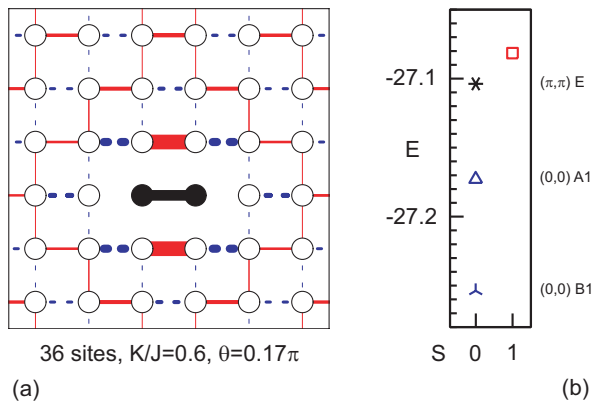}}
  \caption{
    (Color online)
    Staggered Valence Bond Crystal phase: (a) Dimer correlations
    in the groundstate of a 36 site sample at $\theta\approx 0.17\pi$.
    Full, red (dashed, blue) lines denote negative (positive) correlations.
    The line width is proportional to the correlation.
    (b) Low energy spectrum of a 36 sites sample in the same phase.
    The required four singlets (the singlet at $(\pi,\pi)$ marked with a
    star is two-fold degenerate) with the correct quantum numbers
    corresponding to a staggered VBC are found below the first triplet.
    \label{fig:VBC}
  }
\end{figure}
Once the nematic state has been destabilized by even stronger
frustration we find evidence for a VBC state with a staggered dimer
structure.
We consistently see an increase of the staggered dimer structure factor
for all system sizes considered. The real-space dimer correlations for an $N=36$
sample are shown in Fig.~\ref{fig:VBC}a). These correlations show a clear
staggered pattern and they converge to a finite value at the largest distances.
Another strong argument in favor of a staggered dimer phase is the presence of 4
singlets which will form the fourfold degenerate groundstate manifold in the
thermodynamic limit, as shown in Fig.~\ref{fig:VBC}b). We find two distinct
singlets at momentum $(0,0)$ and a doubly degenerate singlet at $(\pi,\pi)$.
The symmetry properties of these singlets precisely correspond to those expected
for a staggered dimer phase. We note that these findings are in line with
earlier work which proposed a staggered dimer phase in a cyclic exchange
model on the square lattice~\cite{cgb92} and recent studies of a two leg
ladder~\cite{lst03,divscalarchiral}.

{\it Conclusion---}
A thorough examination of the low lying spectrum of the
ring exchange Hamiltonian on the square lattice revealed how a
biaxial magnet may be driven to an uniaxial spin-nematic phase
through the interplay of frustration and quantum fluctuations.
These fluctuations drive the disappearance of the net magnetic moment of the
planar orthogonal four-sublattice state, the spins disorder
in the spin plane, the associated Goldstone mode acquires a gap, but
nevertheless in a a finite range of parameters the plane of spins
remains locked. To our knowledge, this is the first clear demonstration
of the existence of such a phase in a two-dimensional quantum magnet.
At last, increasing again the frustration,
the $SU(2)$ symmetry is completely restored, a spin gap opens 
simultaneously with the collapse in
the $S=0$ sector of the four levels leading to the spatial
staggered VBC.
This is one realization of the two-step restoration of symmetry
speculated some years ago by Chandra and Coleman \cite{cc91}.

In some aspects, such as spin susceptibility and thermodynamic properties,
the spin nematic phase is not different from a standard
N\'eel phase \cite{ag84}. But the absence of long range order in ordinary 
spin-spin correlation functions implies that such systems -- although ordered -- 
do not display a Bragg peak with unpolarized neutrons. 
Due to the pseudo-vectorial nature of the order parameter, the NMR pattern of 
the nematic state is different from that of an ordinary N\'eel phase (different
selection rules), providing a clear experimental signature of this phase.
It might, however, be obscured by a phenomenon not addressed
in this paper, which is the sensitivity to disorder and  impurities: disorder might be able to
locally pin the spins in the transverse plane leading to a novel type of spin glass.

\acknowledgments We acknowledge very interesting discussions with
C.~Berthier and G.~Misguich and R.R.P.~Singh. A.L.~acknowledges
support by the Swiss National Fund and the CNRS. Computations were
performed at the LRZ M\"unchen, the CSCS Manno and at IDRIS Orsay.


\begin{thebibliography}{}
\bibitem{NoteCyclic}
 The ring-exchange Hamiltonian used in fitting
 the experiments on La$_2$CuO$_4$ is somewhat
 different from our model Hamiltonian.
 Restricting the comparison to only genuine
 four-spin terms, $K\approx 1/16 J$ in these
 materials.
\end{thebibliography}
\end{document}